\documentclass[11pt]{article}

\usepackage{hyperref}
\pdfoutput=1

\usepackage{natbib}

\begin{document}
\title{Deterioration of Damselfly Flight Performance due to Wing Damage}
\author{Yan Ren, Zhe Ning, Kuo Gai, Chengyu Li, Samane Zeyghami, Haibo Dong\\
        \vspace{6pt} Department of Materials and Mechanical Engineering, \\
        Wright State University, Dayton, OH 45435, USA}
\maketitle
\begin{abstract}
This is a short introduction illustrating movies submitted to "fluid dynamics videos".
\end{abstract}
\section{Introduction}
In this video, effect of chordwise damage on a damselfly (American Rubyspot)'s  wings is investigated. High speed photogrammetry was used to collect the data of damselflies' flight with intact and damaged wings along the wing chord. Different level of deterioration of flight performance can be observed. Further investigation will be on the dynamic and aerodynamic roles of  each wing with and without damage. 
\bibliographystyle{jfm}
\bibliography{bib}

\end{document}